\documentclass[%
 reprint,
 amsmath,amssymb,
 aps,
]{revtex4-2}

\usepackage{graphicx}
\usepackage{dcolumn}
\usepackage{bm}
\usepackage{xcolor}
\usepackage{appendix}

\begin{document}

\preprint{APS/123-QED}

\title{Cryogenic operation of MEMS-based suspended high overtone bulk acoustic wave resonators for microwave to optical signal transduction}

\author{Stefano Valle}
 \email{stefano.valle@bristol.ac.uk}
\author{Krishna C. Balram}%
 \email{krishna.coimbatorebalram@bristol.ac.uk}
\affiliation{%
 Quantum Engineering Technology Labs and Department of Electrical and Electronic Engineering, University of Bristol, Woodland Road, Bristol BS8 1UB United Kingdom\\
}%

\date{\today}

\begin{abstract}
Suspended high-overtone bulk acoustic wave resonators (HBARs) can serve as a viable optomechanical platform for efficient transduction of signals from the microwave to the optical frequency domain. In contrast to 1D nanobeam optomechanical crystals, HBARs benefit from very high RF to phonon injection efficiency ($\eta_{PIE}\approx$ 1) and low optical pump induced heating at cryogenic temperatures. By building small mode volume optical cavities around these devices, one can in principle achieve optomechanical cooperativities comparable to 1D nanobeam optomechanical crystals. In this work, we demonstrate cryogenic operation ($\approx$ 10 K) of such suspended HBAR devices and show classical signal modulation upto 3.5 GHz and response times $\sim$ 524 ns (for the fundamental mode at 340 MHz). While the transduction efficiency is currently limited by the material and device fabrication processes used in this work, we show that with reasonable modifications, efficient quantum  transduction is within reach using this approach.    
\end{abstract}

\maketitle

\section{Introduction}
As quantum technologies move from the era of isolated quantum devices to networked quantum systems, better methods for interconnecting these devices at scale, while preserving quantum coherence become paramount. In particular, superconducting devices are currently the leading candidate for building large-scale quantum systems approaching thousands of qubits in the near future and underpinned Google's pioneering quantum supremacy experiment \cite{arute2019quantum}. It has also been clear that further scaling of these devices requires the development of efficient quantum transducers that can connect distributed superconducting qubit based processors \cite{awschalom2021development}. Given that telecom wavelength photons are a prerequisite for long-distance quantum communication, such quantum transducers need to be able to (bi-directionally) translate quantum states between the 3-10 GHz microwave quantum states originating in the superconducting qubit to the 194 THz optical photons that can be sent down low-loss optical fibers. Quantum microwave to optical signal transduction \cite{zeuthen2020figures} has been a very active area of research over the past decade and there have been a variety of approaches followed to build efficient transducers, with the three main device platforms being electro-optic, acousto-optic and opto-magnonic devices \cite{chu2020perspective,lauk2020perspectives,han2021microwave}. Of the three, acousto-optic or piezo-optomechanical platforms have attracted the most interest, building on the spectacular advancements in the field of cavity optomechanics \cite{aspelmeyer_cavity_2014} in the past decade, culminating in the recent demonstration of transduction of microwave photons from a superconducting qubit to the optical domain \cite{mirhosseini2020superconducting}. Microwave to optical transduction using piezo-optomechanical devices can be thought of as the interaction of two coupled subsystems: an RF sub-system that converts the input microwave field to the acoustic domain and an optomechanical sub-system that converts this acoustic vibration to the optical domain \cite{wu2020microwave}. Such piezo-optomechanical transducers have been demonstrated in a wide variety of material platforms, ranging from aluminum nitride \cite{bochmann_nanomechanical_2013, Vainsencher_Bi_2016, han_cavity_2020}, aluminum nitride on silicon \cite{mirhosseini2020superconducting}, gallium arsenide (GaAs) \cite{balram_coherent_2016, forsch2020microwave}, gallium phosphide \cite{honl2021microwave} and lithium niobate \cite{shao2019microwave,jiang_efficient_2020}. With few exceptions \cite{han_cavity_2020}, the vast majority of the transducers demonstrated so far have relied on using 1D nanobeam optomechanical crystals \cite{chan_optimized_2012} for engineering the optomechanical sub-system on account of their high optomechanical coupling strength ($g_{0}$) originating from confinement of the optical and mechanical modes in wavelength scale cavities. On the other hand, the demonstrated (cryogenic) photon transduction efficiency of current state of the art transducers is $\approx 10^{-5}$ \cite{jiang_efficient_2020, mirhosseini2020superconducting} and it is limited by two key factors. The first is the phonon injection efficiency ($\eta_{PIE}$) which measures the conversion efficiency of the applied RF signal into the desired mechanical mode of the optomechanical cavity. With 1D optomechanical crystals, the RF signal is first converted into a mechanical mode using a focused interdigitated transducer (IDT) which is then launched onto the optomechanical cavity to excite the breathing mode, which has high $g_{0}$. As discussed in detail in \cite{balram2021piezoelectric}, it is very challenging to design focusing transducers that can simultaneously impedance match to 50 $\Omega$ and mode match to the desired mode of the optomechanical cavity. In addition, unlike in integrated photonics, acoustic waveguides are inherently multi-moded which makes it challenging to engineer the desired mode transformations with high efficiency. The second challenge with optomechanical crystals is that the small cavity mode volume and high surface to volume ratios of these nanoscale cavities make them susceptible to optical pump induced heating effects. Given that quantum transduction experiments need to be carried out in cryogenic environments (T $\approx$ 100 mK), pump induced surface heating limits the total number of intracavity pump photons that can be used to boost the transduction efficiency. The key figure of merit for the optomechanical sub-system is the optomechanical cooperativity ($C_{om} \propto g_{0}^2N_{cav}$) which scales linearly with the number of intracavity pump photons ($N_{cav}$). Cryogenic transduction experiments, especially in dilution fridges, are currently limited to pulsed pump operation with an intracavity photon number $<$ 400 \cite{forsch2020microwave,ramp_elimination_2018,mirhosseini2020superconducting, balram2021piezoelectric}.

\begin{figure}[!tbhp]
    \centering
    \includegraphics[width = \columnwidth]{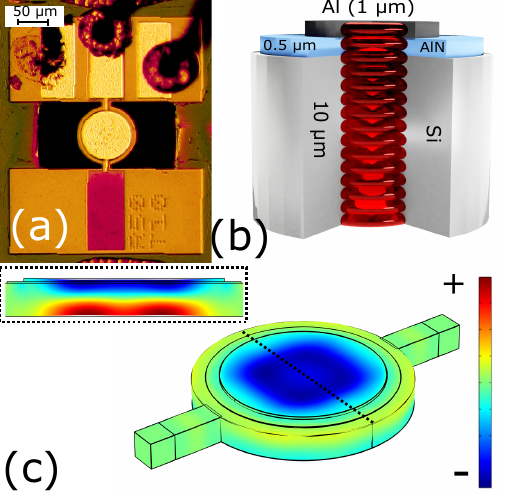}
    \caption{(a) Microscope image of a representative device (disk radius R = 50 ${\mu}m$ corresponds to the radius of the central AlN disk). The signal and ground pads are wire bonded to a PCB for RF interfacing. (b) 3D illustration of the device stack showing the piezoelectric AlN film sandwiched between the top Al and bottom Phosphorous-doped Si electrode. Applying an RF field between the two electrodes generates an acoustic wave that is mostly confined under the top Al electrode. The HBAR resonances correspond to the longitudinal acoustic Fabry-Perot modes of the suspended structure.  (c) 3D FEM simulation of the fundamental ($t_{stack} = {\lambda_a}/2$) HBAR mode of the structure ($f_m$ = 312 MHz). The cross-section displacement is shown in the inset.}
    \label{fig:schematic}
\end{figure}

Given these constraints, it is natural to consider whether efficient quantum transduction is more feasible in macroscopic transducer geometries. High overtone bulk acoustic wave (HBAR) resonators are very attractive in this regard. They can be thought of as quasi-1D acoustic Fabry-Perot cavities and are widely deployed as acoustic wave filters in modern smart phones \cite{campbell_surface_1998,hashimoto2009rf}, and in contrast to the nanobeam devices, provide two key advantages. On the RF sub-system front, the $\eta_{PIE}$ in these devices can approach unity \cite{gokhale2020epitaxial}, as they can be easily impedance matched to 50 $\Omega$, even at very high frequencies ($\sim$ 10 GHz). HBAR devices have already been interfaced with superconducting qubits \cite{chu2017quantum,kervinen2018interfacing} and swapping of quantum states between the microwave and acoustic domains has been demonstrated. By using epitaxial techniques for metal deposition, high frequency acoustic dissipation and scattering can be minimized , leading to cryogenic $f_{m}*Q_{m}$ products that exceed $10^{17}$ at 10 GHz \cite{gokhale2020epitaxial}. In addition, given these devices are by definition quasi-bulk devices and have much lower surface to volume ratio in comparison to the nanobeam cavities, they are relatively immune to pump induced optical heating effects and the optomechanical interaction can be parametrically enhanced by the number of intracavity pump photons ($N_{cav}$). The main limitation of HBAR devices from a quantum transduction perspective is the large cavity mode volume and weak optomechanical interaction strength, which reduces the achievable $g_{0}$ to the 100s Hz range, compared to nanobeam optomechanical crystals, which currently have $g_{0}$ $\approx$ 1 MHz \cite{Balram_Moving_2014}.  The interaction strength in HBAR devices is weak if we rely primarily on the moving boundary effect. If we instead operate at the Brillouin scattering frequency and exploit materials with large elasto-optic ($p_{12}$) coefficients \cite{renninger2018bulk}, the optomechanical interaction is phase matched and the $g_{0}$ can be significantly enhanced. To understand the scaling, it is instructive to consider a quasi 1D interaction between optical and acoustic modes in a Fabry-Perot like cavity, which closely approximates the HBAR scenario considered in this work. In this case, the photoelastic effect dominated optomechanical coupling strength ($g_{0,PE}$) at the Brillouin frequency (phase-matching condition) can be approximated as \cite{renninger2018bulk, valle_high-frequency_2019}: 
\begin{eqnarray}
g_\text{0,PE} \sim \frac{\omega_{c}^2n^3p_{12}}{2c}\sqrt{\frac{\hbar}{2{\rho}A_{eff}L\Omega_{m}}}
\label{eq:g0_pe}
\end{eqnarray}
\begin{figure}[!htb]
    \centering
    \includegraphics[width = \columnwidth]{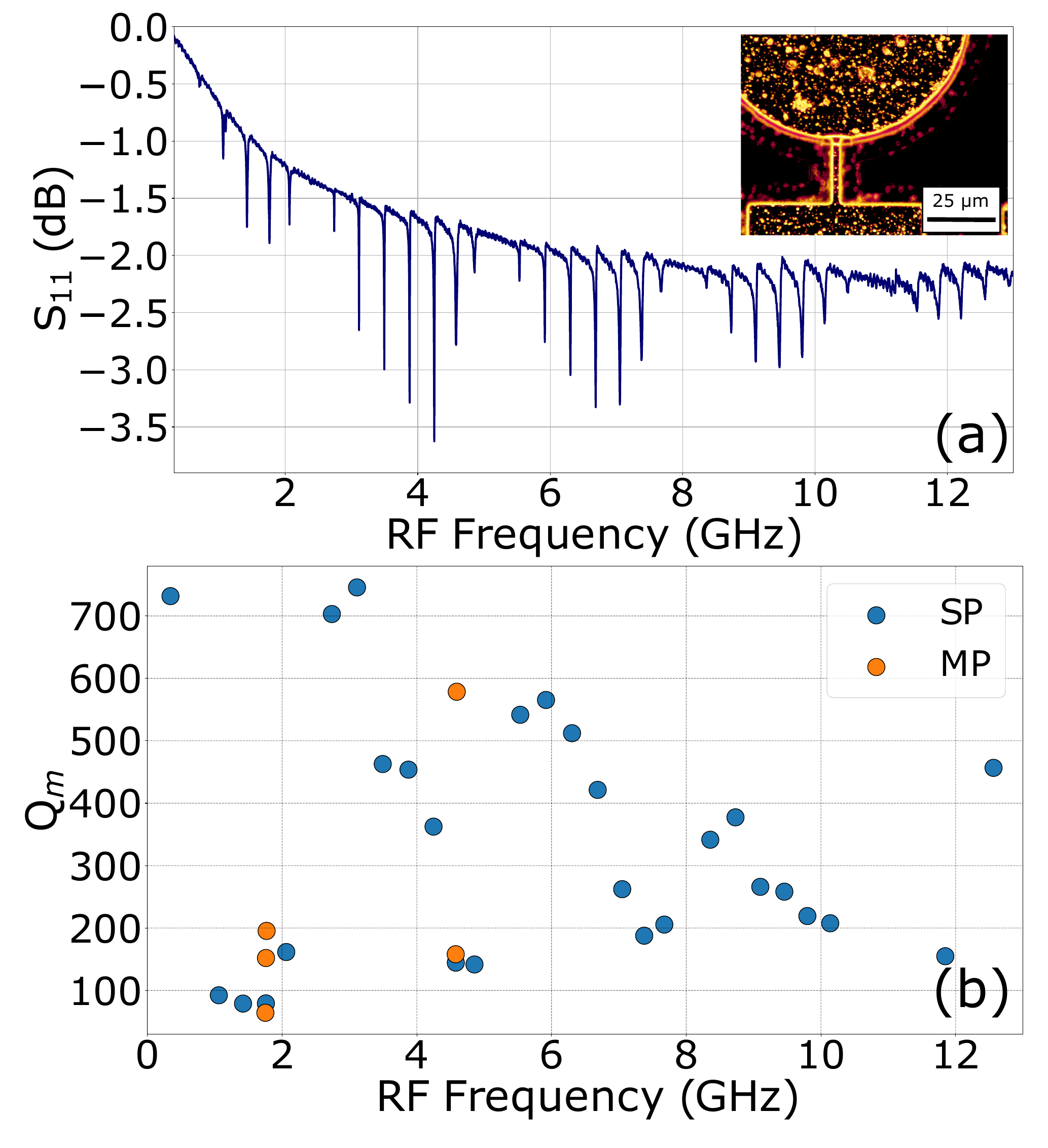}
    \caption{(a) RF reflection ($S_{11}$) spectrum of an R = 50 ${\mu}$m disk measured at room temperature. The HBAR resonances can be easily observed upto 12 GHz. A dark field microscope image of the disk is shown in the inset showing the surface roughness of the Al layer that limits our $Q_{m}$ (b) Extracted quality factor ($Q_{m}$) from a single peak (SP) Lorentzian fit of each of the modes in (a). Due to the symmetry of the electrodes, we see mode splitting in some of the modes and fit the mode with a multi-peak fit (MP). See Appendix \ref{Lorentz Multi-peak fit} for further details.}
    \label{fig:S11_HF}
\end{figure}
where $\omega_{c}/2\pi$ is the optical cavity frequency (Hz), $n$ is the refractive index of the medium, $p_{12}$ is the photoelastic coefficient mediating the interaction between a tranvserse electric (TE) polarized optical wave and a longitudinal acoustic wave, $c$ is the speed of light in vacuum, $\rho$ is the material density, $A_{eff}$ is the interaction cross-section area between the acoustic and optical fields, $L$ is the cavity length and $\Omega_m/2\pi$ is the mechanical mode (Brillouin) frequency (Hz). As eq.\ref{eq:g0_pe} shows, $g_{0}\propto\sqrt{\frac{1}{A_{eff}L}}$ which means that small mode volume ($V_{m}\propto A_{eff}L$) HBAR cavities can significantly enhance the optomechanical interaction. Traditional HBARs \cite{hashimoto2009rf} are built on a low-loss acoustic substrate like silicon or sapphire with a substrate thickness $\approx$ 500 ${\mu}m$. By moving to a MEMS derived fabrication process and working with a suspended HBAR instead, the substrate thickness can be reduced to the 5-10 ${\mu}m$ range, while preserving the acoustic quality factor ($Q_{m}$). A second challenge with HBAR approaches, as discussed above, is that the interaction strength is strongly peaked at the Brillouin scattering frequency which is primarily determined by material properties. In a simple 1D interaction, the Brillouin frquency ($f_{B}$) can be written as $f_B = \frac{2nv_{a}}{\lambda_{p}}$ where $n$ is the refractive index of the medium, $v_{a}$ is the speed of sound and $\lambda_{p}$ is the optical pump wavelength. Traditional HBAR substrates like silicon and sapphire have fast acoustic velocities which pushes the Brillouin frequency far beyond the 3-10 GHz range in which most state-of-the art superconducting qubits currently operate. While the current work uses Si based HBAR devices, primarily because of ease of fabrication through a commercial foundry platform, in section \ref{quantum_hBAR} we show how the designs presented here can be modified to achieve quantum transduction in a suspended HBAR platform.
\section{Device design and operation}
Fig.\ref{fig:schematic}(a) shows a microscope image of a representative HBAR device, fabricated using the PiezoMUMPS foundry platform \cite{valle_high-frequency_2019, cowen2014piezomumps}. The device (cross-section indicated in Fig.\ref{fig:schematic}(b) consists of a 10 ${\mu}m$ thick Phosphorus-doped suspended silicon layer, on which 500 nm of piezoelectric aluminum nitride (AlN) is deposited with a top electrode of 1 ${\mu}$m aluminum (Al). By applying an RF field between the top Al electrode and the bottom doped Si layer, the piezoelectric AlN layer can be set into vibration. At certain frequencies (corresponding to the 1D acoustic Fabry-Perot resonant modes of the structure), the whole suspended stack (Al + AlN + Si) is set into coherent vibration. Since most of the elastic energy primarily resides in the low-loss silicon layer, such HBAR modes can achieve very low loss and high mechanical quality factors ($Q_{m}$), provided that roughness induced scattering can be sufficiently controlled \cite{galliou2013extremely, gokhale2020epitaxial}. HBAR devices are generally well-suited for achieving high $Q_m$ at ultra-high frequencies ($\approx$ 10 GHz). Especially in comparison to surface acoustic wave (SAW) devices, where higher frequencies result in higher insertion loss mainly due to reducing finger widths in interdigitated transducers \cite{datta1986surface}, HBAR devices can achieve very low insertion loss and maintain high $Q_{m}$ beyond 5 GHz \cite{hashimoto2009rf}. One way to look at this is to see that while SAW filters dominate RF filters at the $<$ 1 GHz frequency range, all modern RF filters beyond 2 GHz are BAW based. In addition to higher $Q_{m}$, HBAR devices have another significant advantage, which is critical for quantum transduction. HBAR devices usually have a phonon injection efficiency ($\eta_{PIE}$) of $\sim$ 1. $\eta_{PIE}$ is defined as the fraction of the input RF energy that is converted to acoustic energy in the mode of interest. For comparison, in SAW devices while it is feasible to convert all the RF energy into propagating SAW waves with suitable impedance matching \cite{wu2020microwave}, it is challenging to excite the mode of interest (usually the breathing mode of a 1D optomechanical crystal) with very high efficiency using this approach \cite{balram2021piezoelectric}. This results in a low overall $\eta_{PIE}$. In comparison, by building a quantum transducer around one of the HBAR modes, as we discuss in section VI, we are working with a system in which $\eta_{PIE}\approx$ 1 since almost all the incident RF energy is transferred to a mechanical mode with high $g_0$..

Fig. \ref{fig:S11_HF}(a) shows the measured HBAR modes of a representative device (with R = 50 $\mu m$, where the radius corresponds to the size of the piezoelectric AlN film. The Al and Si disks have a size of R-5 ${\mu}$m and R+5 ${\mu}$m respectively). The modes are clearly evident in the RF reflection ($S_{11}$) spectrum of the device. At ambient conditions, we are able to observe HBAR modes up to 12 GHz. As discussed above, these HBAR modes are primarily thickness modes of the stack and the modal displacement profile of the fundamental mode (at $f_m\approx$ 340 MHz), calculated using COMSOL MultiPhysics$^\text{TM}$ is shown for reference in Fig.\ref{fig:schematic}(c).  Due to the mass loading from the top Al electrode, the higher frequency modes are, in general, more confined towards the centre of the disk away from the tethers \cite{valle_high-frequency_2019}. The quality factor ($Q_m$) of each mode, as extracted from a Lorentzian fit to the spectrum is shown in Fig. \ref{fig:S11_HF}(b). We believe the scatter in the $Q_{m}$ data comes from two main sources: the roughness of the top Al electrode (shown in the inset of Fig. \ref{fig:S11_HF}(a) which affects the modes differently)  and the symmetry-breaking induced by the roughness and the presence of the tethers, which results in mode-splitting and intermodal coupling, making it challenging to model the lineshape with a simple Lorentzian (a more detailed analysis is reported in Appendix \ref{Lorentz Multi-peak fit}).\\
\begin{figure}[!hbtp]
    \centering
    \includegraphics[width = \columnwidth]{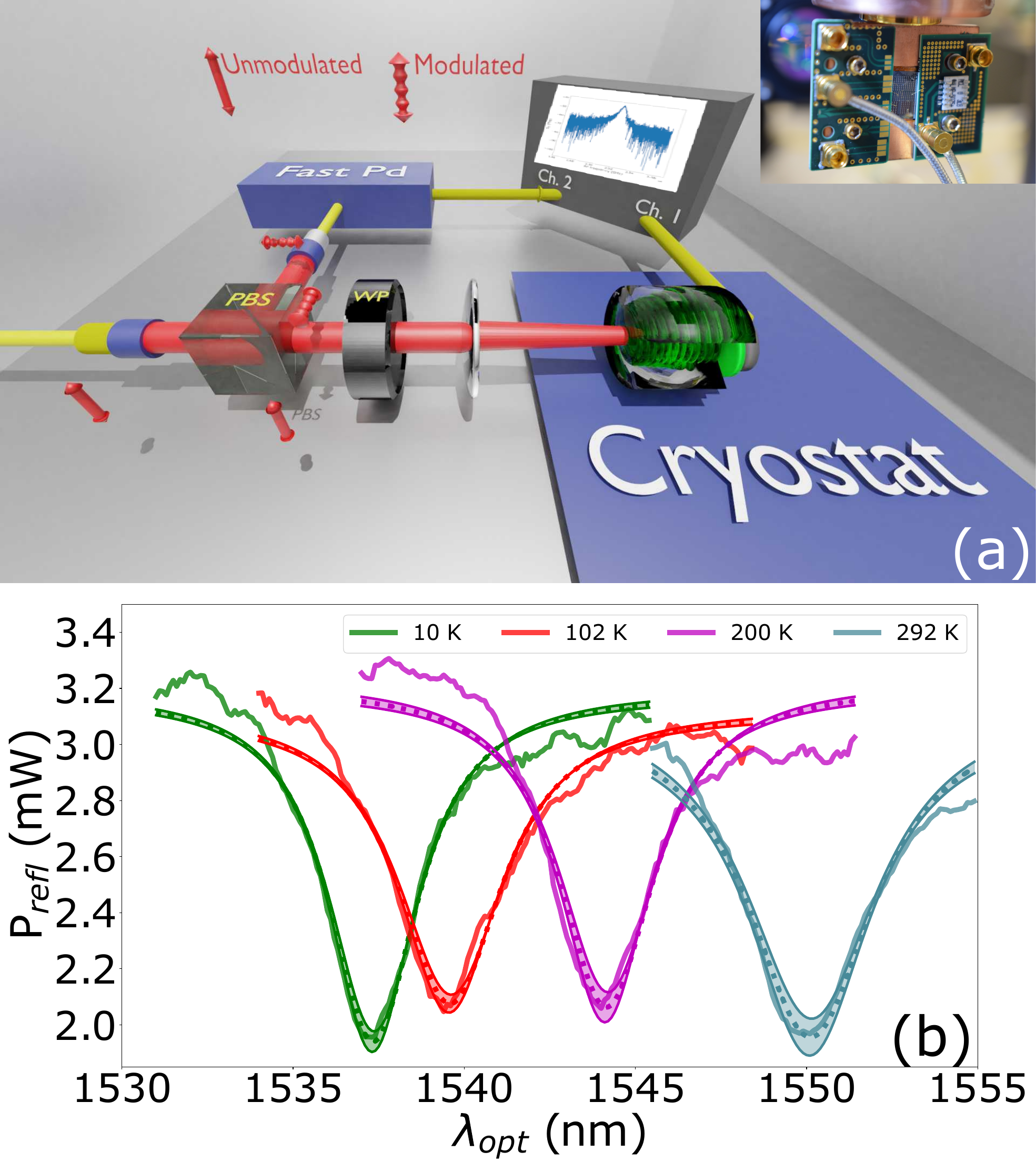}
    \caption{(a) Schematic illustration of the measurement setup used to probe the acousto-optic modulation induced by the HBAR modes. A standard reflection measurement setup using a polarizing beam splitter (PBS) and a quarter wave plate (WP) is built around a closed cycle optical cryostat. The disk is driven from channel 1 of the VNA and output of the fast photodiode is fed into channel 2 of the VNA realizing a coherent acousto-optic $S_{21}$ measurement (see text for details). The polarization state at each step is indicated by red arrows. An image of the HBAR chip on a PCB mounted on the cold finger of the cryostat is shown in the inset. The RF connections are made using SMB connectors (a cable is shown for reference). (c) Optical cavity spectrum as the cavity is cooled down from room temperature to $\approx$ 10 K. The spectra are fit by a Lorentzian lineshape (shown by solid lines). The variation in $Q_{o}$ and the off-resonance reflected power is due to a change in the system alignment during the cooldown process.}
    \label{fig:Setup}
\end{figure}

The measured $Q_{m}$ in the frequencies of interest for quantum transduction (3-10 GHz), range between $\approx$ 100-750 depending on the mode. We expect to increase $Q_{m}$ by 2-3 orders of magnitude by reducing the surface roughness with an improved metal deposition process  \cite{mcpeak2015plasmonic},  especially at cryogenic temperatures.

\section{Cryogenic high-frequency acousto-optic modulation}

The devices are mounted on a copper sample holder with a custom RF PCB attached (shown in the inset of Fig. \ref{fig:Setup}(a)), which allows the routing of the RF signal via a rigid SMA RF cable and wirebonds from the RF source to the device under test. A preliminary room temperature characterization is carried out to verify device operation before moving to the cryogenic measurements. The sample holder is mounted on the cold finger of a closed-cycle optical cryostat (ARS) with a nominal base temperature of 7 K. The minimum temperature achieved in our experiments is $\approx$ 10 K, due to the thermal load exerted by the RF cable and feed-through (at room temperature) connected to the cryogenic sample stage. With better thermal shielding, we expect to be able to achieve the nominal base temperature (7 K) in future experiments. The temperature of the device is monitored by using a temperature sensor (Lakeshore DT-621-HR) connected to the cold finger. In our experiments, we assume that the sample is sufficiently thermalized and its temperature is identical to the temperature of the cold finger, as read by the sensor.\\
The acousto-optic characterization of the device (R = 35 ${\mu}m$) is performed using a setup similar to our previous work \cite{valle_high-frequency_2019} and shown in Fig. \ref{fig:Setup}(a). The device is operated in reflection from the silicon side, which is accessible in this foundry process (see black windows in Fig. 1(a)). By adding a thin  layer (20-25 nm) of gold (Au) to the back side of the suspended silicon, we can improve $Q_{o}$ by 2-3$x$ to $\sim$ 450 in our experiment. Light from a tunable telecom wavelength laser (Santec TSL550) is circularly polarized using a quarter-wave ($\lambda/4$) plate, and focused on the device. The reflected light is separated from the incident beam using a polarizing beam splitter and is detected using a high speed photodiode (Thorlabs DET08 CFC/M). A vector network analyzer (Rohde$\&$Schwarz ZVL) is used to drive the device (Port 1) and the detected signal from the photodiode is fed into Port 2 to measure the coherent acousto-optic modulation signal (AO $S_{21}$). 

Fig. \ref{fig:Setup}(b) shows a plot of the measured optical cavity reflectivity as a function of temperature. The incident spot size is estimated to be $\approx$ 70 ${\mu}m$ (beam diameter) and the incident optical power is 5 mW. As shown in Fig. \ref{fig:Setup}(b), the temperature dependence of silicon's refractive index shifts the optical cavity spectrum and by fitting a 1D Fabry-Perot model, we can get an independent estimate of the sample temperature and compare it with the reading on the cold finger. We would like to note here that for the low $Q_{o}$ cavities used in this work, this method only works accurately for temperatures $>$ 30 K as silicon's thermo-optic coefficient is reduced considerably below that. All temperatures below 30 K stated in this work are readings from the cold finger, but we believe our sample temperatures are very close to this value. In addition to a change in the optical cavity frequency, Fig. \ref{fig:Setup}(b) also shows an apparent change in $Q_{o}$. This is mainly an artefact of the need to re-align the experiment during the sample cooldown when these spectra were collected. 

\begin{figure}[!hbtp]
    \centering
    \includegraphics[width = \columnwidth]{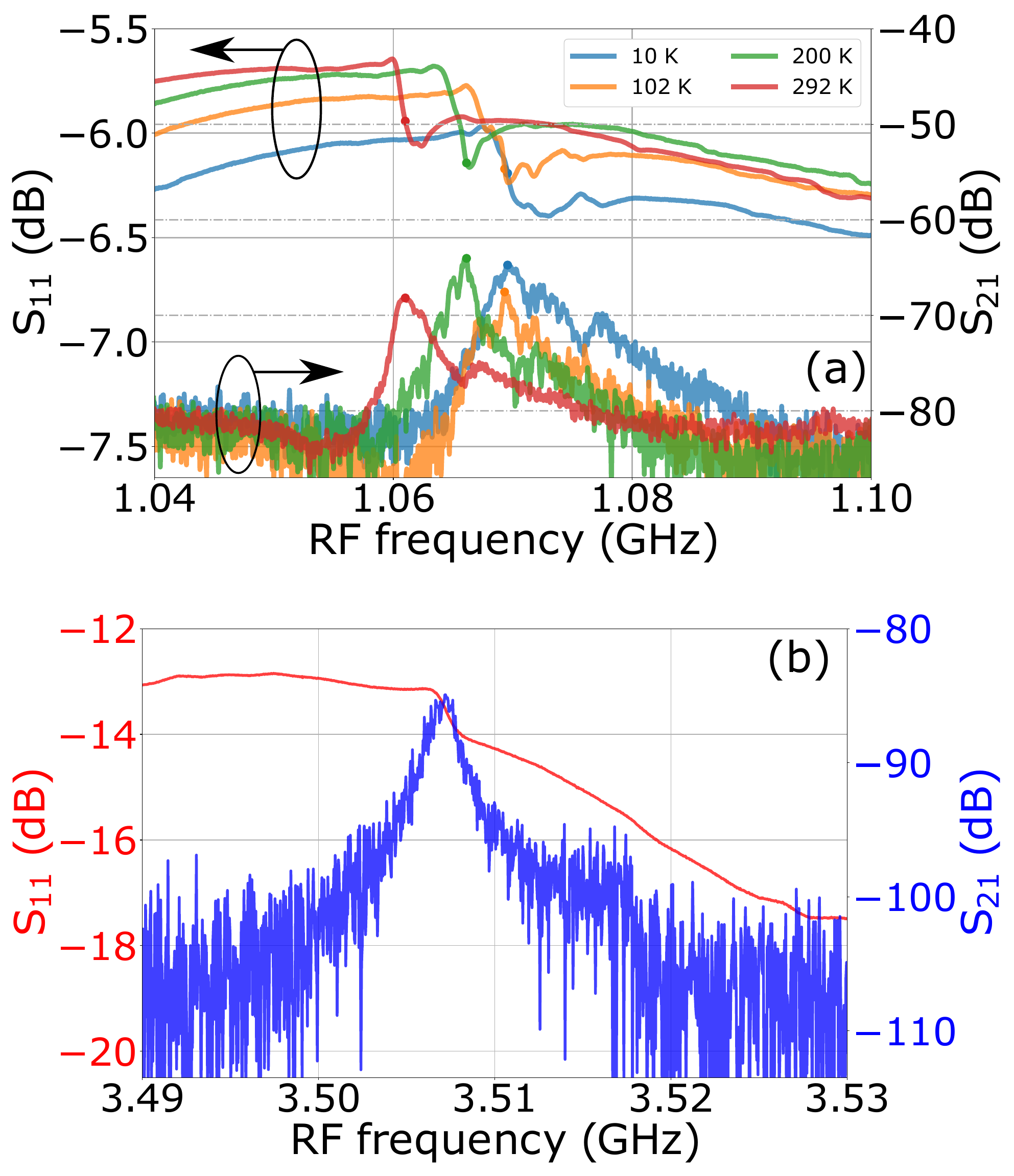}
    \caption{(a) RF reflection ($S_{11}$) and acousto-optic transmission ($S_{21}$) spectrum of the $\approx$ 1 GHz HBAR mode of an R = 35 ${\mu}m$ device at various temperatures ranging from room temperature (292 K) to 10 K. The RF power entering the cryostat is 1 mW and the optical power is 5 mW. At low temperatures, the 1 GHz resonance shows a multi-peak behavior (seen in both the $S_{11}$ and $S_{21}$ spectra) that we don't fully understand, but attribute mainly to the stress induced during cooldown. (b) AO $S_{21}$ spectrum of the highest  frequency ($\approx$ 3.5 GHz ) HBAR mode that we could observe in our experiment at 10 K with 10 mW of optical input power.}
    \label{fig:S21-S11}
\end{figure}

We can also observe the HBAR resonances of the cavity at cryogenic temperatures in the $S_{11}$ spectrum (Fig. \ref{fig:S21-S11}(b)). In contrast to the room temperature data shown in Fig. \ref{fig:S11_HF}(a) recorded using an RF probe with calibrated response, the HBAR modes in the cryogenic $S_{11}$ spectrum are not distinctly visible due to the challenge of de-embedding the phase error introduced by the PCBs and wirebonds as shown by the $S_{11}$ spectrum in Fig. \ref{fig:S21-S11}(a). On the other hand, the modulation peaks in the cryogenic AO $S_{21}$ spectra correspond exactly to the HBAR modes of the suspended opto-mechanical cavity and show similar spectral signatures, (esp. at high frequencies $>$ 2 GHz, see Fig. \ref{fig:S21-S11}(b)) to the room temperature measurements.  As discussed in our previous work \cite{valle_high-frequency_2019}, the modulation is dominated primarily by the moving boundary effect \cite{Balram_Moving_2014} since the cavity length periodically expands and contracts on the excitation of the HBAR mode, which changes the optical cavity frequency. This effect is significantly enhanced in the suspended platform with both cavity boundaries free to move. In our experiments (with incident optical power 10 mW), we are able to observe cryogenic ($\approx$ 10 K) modulation up to $\approx$ 3.5 GHz (shown in Fig. \ref{fig:S21-S11}(b)). 

In these experiments, the device performance (photon transduction efficiency) is limited primarily by the $Q_{o}$, $Q_{m}$ and the RF delivery efficiency inside the optical cryostat. In contrast to traditional piezo-optomechanical platforms which have relied on small mode-volume and high $Q_{m} \approx 10^4, Q_{o} \approx 10^5$ 1D nanobeam optomechanical crystals \cite{balram_coherent_2016, stockill2019gallium, mirhosseini2020superconducting, jiang_efficient_2020} with high optomechanical coupling strengths ($g_{0}\approx 1$ MHz), our current devices have $Q_{o} <$ 500, $Q_{m} < 10^3$ , and $g_{0}\approx$ 2$\pi$ * 642 Hz (for the fundamental mode at 340 MHz, $R$ = 35 $\mu$m). In spite of these limitations, we are still able to observe cryogenic transduction \cite{forsch2020microwave} with moderate photon conversion efficiencies  $\approx$ 3.47*$10^{-10}$ (uncorrected) and $\approx 4.43*10^{-8}$ accounting for the actual RF power (8 ${\mu}$W for 1 mW  input) being delivered to the transducer and 10 mW optical pump power (see Appendix \ref{modulation index} for details). It is important to note that the low $Q_{o}$ results in approximately equal transduction into both upper and lower sidebands, so the intrinsic transduction efficieny is a factor of 2 higher ($\approx 10^{-7}$).

\begin{figure}[!hbtp]
    \centering
    \includegraphics[width = 0.75\columnwidth]{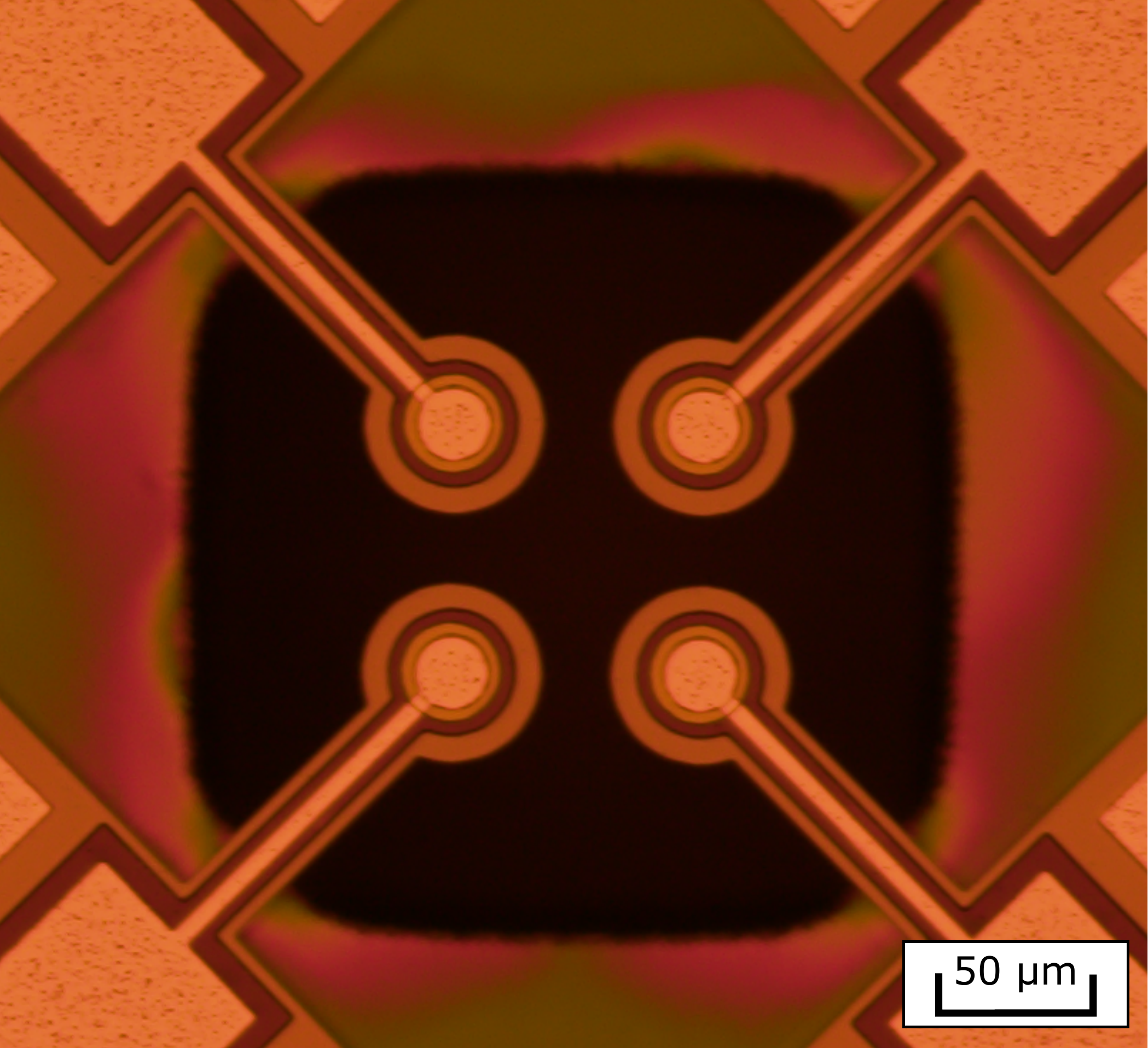}
    \caption{Microscope image of a 2x2 multi-pixel HBAR AOM device with R = 15 $\mu$m, as a prototype building block of AO cryogenic spatial light modulators.}
    \label{fig:lattice}
\end{figure}

While the overall transduction efficiency is understandably poor in these devices, we show in section \ref{quantum_hBAR} that all of these parameters ($Q_{m}$, $Q_{o}$ and $g_{0}$) can be significantly improved to achieve quantum transduction in this geometry. The main reasons for our optimism in the suspended HBAR platform for quantum transduction, despite the sub-optimal device parameters are outlined here: we can parametrically enhance the interaction strength (and the modulation efficiency) by increasing the intracavity optical pump power (as shown in Appendix \ref{Optical power dep}). Given the relatively macroscopic size of these devices, they are not susceptible to the optical pump induced surface heating effects, which have limited nanobeam cavities to pulsed operation and low intracavity photon numbers ($N_{cav}$) in dilution fridge experiments \cite{balram2021piezoelectric}. In contrast to the nanobeam systems which have $N_{cav} < $ 500, we can readily achieve $N_{cav} \approx 10^5$ in our experiments at 11 K. Given the surface to volume ratio of these devices, we don't expect $N_{cav}$ to change significantly as $Q_{o}$ is improved in future devices or the device is operated at much lower temperatures, although we haven't confirmed this experimentally. The other advantage of these HBAR approaches is that $\eta_{PIE}\approx$ 1 and the system can be effectively impedance matched and have low RF insertion loss ($< 0.2$ dB in state-of-the art RF filters \cite{hashimoto2009rf}). For 1D nanobeam cavities, the current $\eta_{PIE} < 10^{-3}$ and requires sophisticated impedance matching to push the efficiency higher \cite{wu2020microwave}. The HBAR devices avoid both of those issues in practice. The low $g_{0}$ that limits the transduction efficiency (and operating frequency) in these experiments is primarily due to relying on the moving boundary effect. As we show in section \ref{quantum_hBAR}, this can be addressed by moving towards the Brillouin scattering condition and exploiting the much stronger electrostrictive coupling.

\section{Pulsed operation}

In addition to their use as microwave to optical signal transducers, one of the advantages of the suspended HBAR devices is their use in building a 2D array of cryogenic acousto-optic modulators for addressing trapped-ion and neutal atom based quantum information platforms \cite{wright2019benchmarking}. As the number of qubits in these platforms starts to grow, optical beam addressing presents a significant challenge to system scaling. A fast (sub-${\mu}$s response time), low-power and scalable (microfabricated) AOM platform with large pixel count is required to address these issues. A representative device fabricated in the same platform as a 2x2 array is shown in Fig. \ref{fig:lattice}. Given that the current HBAR devices have moderate $Q_{o}$ and $Q_{m}$ and depend on film thickness for both their optical and mechanical resonances, it is feasible to construct a large 2D array of these modulators without requiring individual tuning elements on each. To explore their use as building blocks for future cryogenic AOM arrays, \cite{balram_acousto-optic_2017}, we characterized the cryogenic performance of the device in the time domain using pulsed RF inputs.

\begin{figure}[!htbp]
    \centering
    \includegraphics[width = \columnwidth]{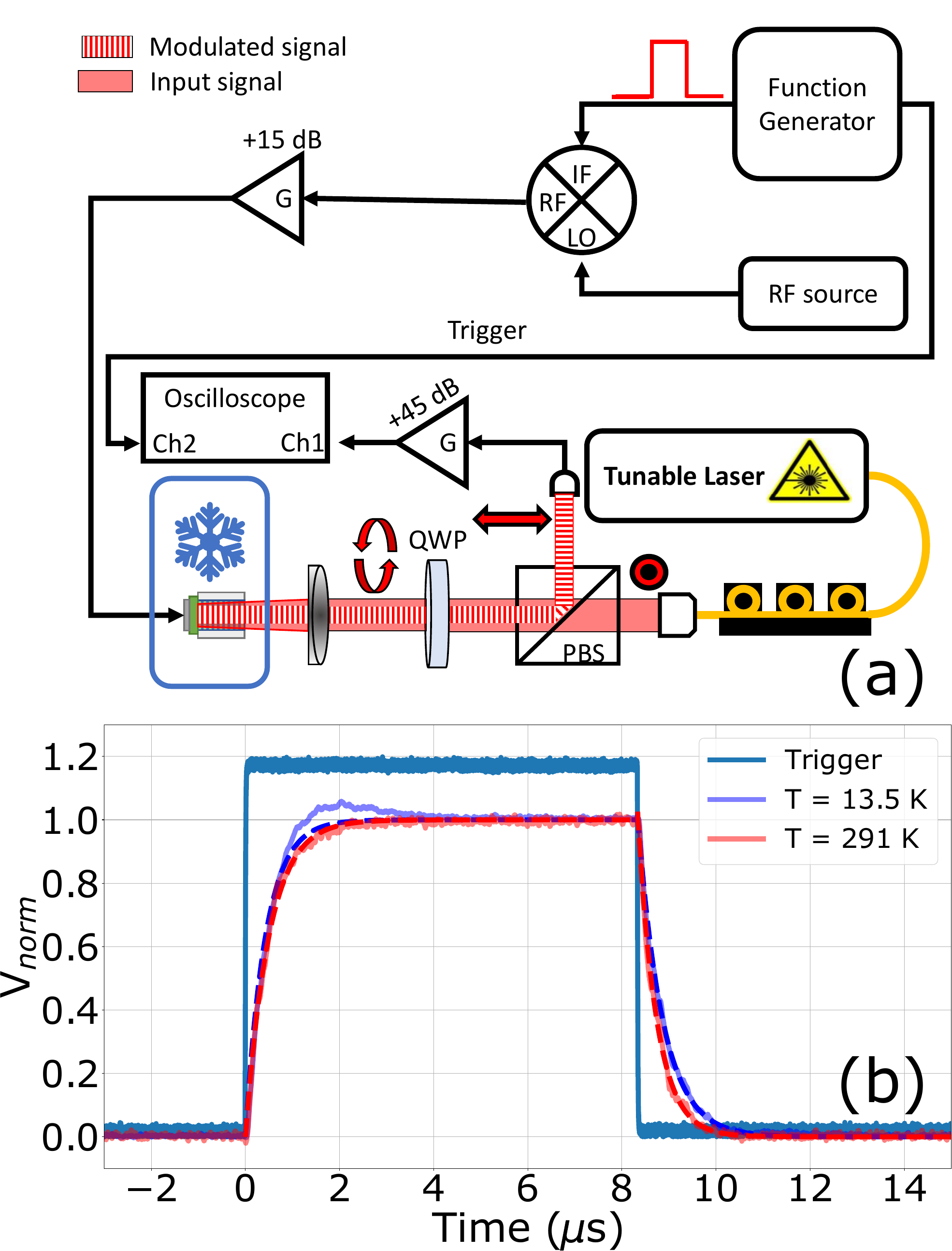}
    \caption{(a) Schematic of the measurement setup used to characterize the time domain response of the suspended HBAR devices, see text for details (b) Measured response of the fundamental 345 MHz mode (R = 35 ${\mu}$m) at room temperature (red) and 13.5 K (purple. The dashed black line is a fit to the cryogenic response. The pulse trigger is shown for reference in blue. The cryogenic response time (524 ns) is extracted from the fall time.}
    \label{fig:mixer_setup}
\end{figure}

Fig. \ref{fig:mixer_setup}(a) shows the measurement setup used for the pulsed measurement. We use an RF mixer (MiniCircuits ZFM-4S+) to generate a pulsed RF signal with a carrier frequency 343.2 MHz at 291 K, and 344.950 MHz at 13.5 K with a pulse repetition rate at 60 kHz and a duty cycle of 50$\%$. The signal is delivered to the device in the cryostat and the reflected (modulated) optical signal is detected using a high speed photodiode and amplified by a cascade of three RF amplifier (MiniCircuits ZX60V62+) with individual nominal gain of +15 dB (net gain $\approx$ 45 dB) before being sent to a high-speed oscilloscope (Keysight Infinium S-Series). The oscilloscope is triggered by the pulse generator. Given the signal to noise in this direct detection experiment is much lower than in the coherent (AO $S_{21}$) experiments in Fig. 2 and 3, we chose to primarily work with the fundamental ($\approx$ 343 MHz) thickness mode of the cavity which shows the largest modulation amplitude \cite{valle_high-frequency_2019} on account of the moving boundary effect ($g_0 \sim g_{0,mb} \approx 2\pi$*642 Hz, for R = 35 $\mu$m).

Fig. \ref{fig:mixer_setup}(b) shows the measured optical response to pulsed input. To characterize the time domain response, we mainly fit the fall time of the system because the rise time at cryogenic temperatures shows some overshoot whose origin we don't fully understand. The fall time at room temperature (291 K) is $\approx$ 398.9 ns at 343.2 MHz, which agrees well with our measured $Q_{m}$ of 950, extracted from RF $S_{11}$ measurements (see Appendix \ref{Lorentz Multi-peak fit}). After cooling down the device to 13.5 K (the base temperature for these experiments is higher than for the AO $S_{21}$ because of the additional RF heating from the amplified input signal), the fall time is equal to 524.5 ns at 344.95 MHz (due to a shift in the mechanical frequency with temperature). The increase in fall time shows an increase of $Q_m$ by $\approx$ 1.3$x$, although the improvement is far lower than expected \cite{gokhale2020epitaxial} because of the excess scattering induced by the surface roughness of the top Al electrode. While the operation speed of these devices is certainly compatible with the gate times (sub-$\mu s$) for trapped ion and neutral atom systems and they satisfy the form-factor and power dissipation (sub-mW RF power/pixel) metrics at cryogenic temperatures, there are still a few challenges that need to be addressed. In particular the pump extinction between the ON and OFF states needs to be significantly improved for these devices to replace traditional bulk AO devices in trapped ion systems. The same factors that limit our transduction efficiency (discussed in more detail in section \ref{Discussion}), in particular, low $Q_{o}, Q_{m}$ and $g_{o}$ limit our overall extinction. The transduction efficiency \cite{valle_high-frequency_2019} (and the extinction) scales $\propto\beta\Omega_{m}/\kappa$ where $\beta$ is the modulation index, $\Omega_m$ is the mechanical mode frequency and $\kappa$ is the optical cavity linewidth. For a fixed $\Omega_{m}$ and RF power (which caps $\beta$), the main avenue for improved extinction is reducing $\kappa$, which is discussed in Section \ref{quantum_hBAR}. It is important to note that, in contrast to traditional AOMs, the collinear operation of these resonant devices, which provides a gain in operation efficiency and scalability, removes the spatial separation between pump and sideband, which precludes spatial filtering methods that is used to boost the extinction in traditional AOMs.  

\section{Discussion: Limitations on transduction efficiency} \label{Discussion}

We estimate the microwave to optical photon transduction efficiency in our cryogenic experiments to be $\approx$ $4.43*10^{-8}$ for the fundamental mode at $\approx$ 342 MHz and an input optical pump power of 10 mW \cite{valle_high-frequency_2019}. Given the challenges of de-embedding the cryogenic $S_{11}$ response as discussed above, to estimate the acoustic energy in the cavity for a given RF power, we use the linear dependence of the acousto-optic $S_{21}$ with RF power. We assume that both $Q_{m}$ and $Q_{o}$ do not change significantly at cryogenic temperatures use the device performance at room temperature (measured using calibrated RF probes) and use it to bound the cryogenic RF transduction efficiency. 

While $\eta_{PIE}$ governs the efficiency of RF-acoustics conversion and is very high in these suspended HBAR geometries, once the acoustic mode of the optomechanical cavity is excited, the optomechanical transduction efficiency is governed by the optomechanical cooperativity  $C_{om}=\frac{4g^{2}_{0}N_{cav}}{\kappa\gamma}$ where $g_{0}$ is the optomechanical coupling strength, $N_{cav}$ is the intracavity photon number, $\kappa$ is the decay rate of the optical cavity mode and $\gamma$ is the mechanical mode decay rate. While $N_{cav}$ in our experiments is large ($\approx 9.33*10^4$ at 10 mW input power and $Q_o \approx$ 460, with the pump laser parked at the point of maximum slope, resulting in an effective coupling ($\eta_{c}$) of 0.5), the other cavity parameters severely limit the transduction efficiency in this experiment. By depositing a gold layer, we can improve $Q_{o}$ slightly, but the flat-flat cavity geometry and the choice of metals for both mirrors  limits the ultimate $Q_{o}$ in this system.

The $Q_{m}$ in these devices is limited currently by surface roughness in the top Al electrode. The inset in Fig. \ref{fig:S11_HF}(a) shows a microscope image of a representative device in dark field clearly illustrating the scale and extent of the roughness. Given the acoustic wave reflects off the top Al boundary, this severely limits the $Q_{m}$ in our devices to be $< 10^3$. We believe the main reason for the surface roughness is the need to deposit thick ($\sim$ 1 ${\mu}$m) films in this MEMS foundry process. As has been demonstrated \cite{gokhale2020epitaxial}, by controlling the surface roughness of metallic films, the $Q_{m}$ of these devices can be significantly enhanced, exceeding $10^6$ at cryogenic temperatures.

The final challenge with these HBAR devices is the low $g_0$, especially as we move to higher frequencies ($>$ 5 GHz) relevant for quantum transduction from superconducting qubits. The acousto-optic interaction in these devices is primarily mediated by the moving boundary effect and the cavity boundary displacement is significantly reduced at higher frequencies. The much stronger Brillouin interactions are only achieved for mechanical frequencies $\sim$ 35 GHz in these structures \cite{valle_high-frequency_2019}. For our current designs (disk radius R = 35 ${\mu}$m), using $g_{om}= 2\omega_c/L_{cav}$ and the zero point motion for the 4 GHz (340 MHz) mechanical mode to be 4.83 (16.6) am for an effective motional mass ($m_{eff}$) $\sim$ 90 pg, we estimate $g_{0} \sim 2 {\pi}$*187 (642) Hz . For the highest $N_{cav}\sim 9.34*10^4$ in our cryogenic experiments and $Q_{m}\approx 600$ $(950)$ for the two modes, we estimate a $C_{om}\approx  4.67*10^{-9}(1.02*10^{-6})$ which is far from the threshold ($C_{om}\approx$ 1) needed for achieving efficient transduction \cite{wu2020microwave}. As discussed in the following section, all of these limitations can be overcome by a suitable choice of materials and device geometries.  

\section{Towards quantum transduction using suspended shear mode HBARs} \label{quantum_hBAR}

\begin{figure}
\begin{center}
\includegraphics[width=\columnwidth]{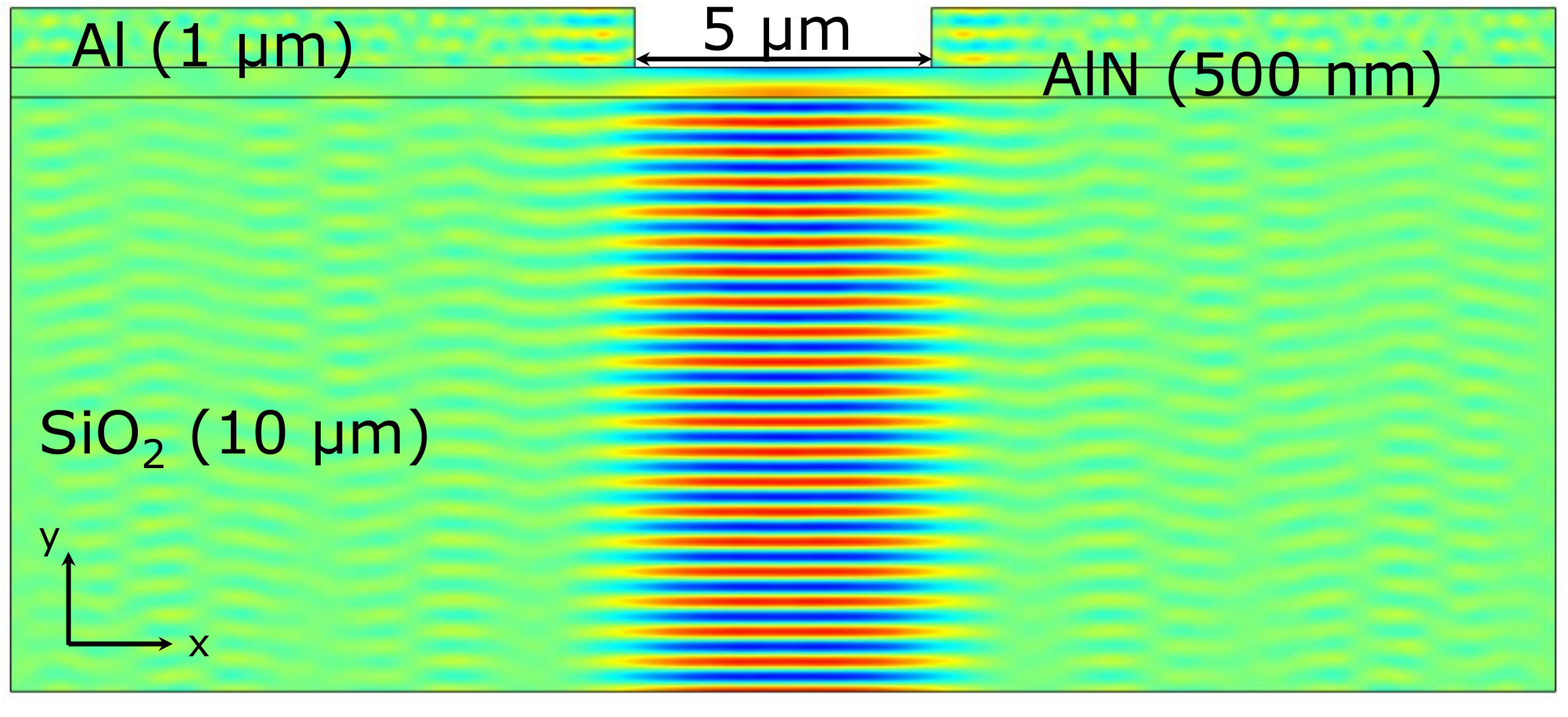}
\end{center}
\caption{ Simulated displacement ($u_x$) of a lateral field excited shear mode resonance in a suspended silicon dioxide resonator ($t_{ox}= 10 {\mu}$m). A 500 nm aluminum nitride (AlN) film is deposited on top of the oxide to excite the resonant mode (resonance frequency $\approx$ 7.3 GHz). This mode satisfies the Brillouin scattering condition for strong phase-matched optomechanical interaction ($g_0\approx 2{\pi}*$ 6.5 kHz)  By building a small mode-volume, high $Q_o$ optical cavity around this device using curved fiber cavities, efficient microwave to optical signal transduction can be engineered in this platform.}
\label{fig:shear_mode_oxide}
\end{figure}

As discussed above, achieving high $Q_m$, $Q_o$ and $g_0$ with the traditional suspended HBAR geometry simultaneously is challenging. On the other hand, all of these issues can be addressed by moving towards a shear wave design. Fig. \ref{fig:shear_mode_oxide} shows a schematic of our proposed structure consisting of 500 nm AlN piezoelectric film deposited on 10 ${\mu}m$ silicon dioxide (SiO$_2$). Instead of working with longitudinal acoustic modes, we instead work with a shear wave resonance that is excited by a lateral field generated by the electrodes. This involves using the $e_{15}$ piezoelectric coefficient of AlN to excite the shear wave, rather than the stronger $e_{33}$ coefficient used in the previous experiments. On the other hand, this shear wave geometry allows us to address all the other issues with suspended HBARs as we discuss below.

\begin{itemize}
    \item $g_{0}$ and $\Omega_{m}$: Shear waves are generally slower than longitudinal waves and hence they allow us to reduce the Brillouin frequency for engineering strong optomechanical interactions in a given material platform. Moreover, the refractive index of silica is $\approx$ 1.46, which allows us to engineer a strong Brillouin interaction (and consequently high $g_0$) at $\Omega_{m}\approx$ 7.3 GHz. This is in contrast to the traditional silicon HBARs, where the Brillouin frequency is $\approx$ 35 GHz (depending on silicon's crystalline orientation). The shear mode resonance in this suspended structure is shown in Fig. \ref{fig:shear_mode_oxide}. The displacement is primarily confined between the electrodes for shear modes, which allows one to achieve much smaller mode volumes as compared to traditional HBARs, a critical requirement for achieving high $g_0$ in these structures. 
    \item $Q_{o}$: The other key advantage that shear modes provide is that the displacement, by being localized between electrodes rather than underneath one (as in traditional HBARs), allows one to engineer much higher $Q_o$ by working with distributed Bragg reflectors for both mirrors. The shear mode silica HBAR can be built on a DBR stack consisting of amorphous silicon and silicon dioxide, with a thin ($\sim$ 500 nm) sacrificial layer (e.g. silicon nitride) for suspending the resonator. Using a similar DBR stack deposited on a 10 ${\mu}m$ radius optical fiber, one can achieve $Q_o > $ 60,000 with an optical cavity mode volume $\approx 63\lambda^3$, assuming a fiber HBAR spacing of $\approx$ 1 ${\mu}m$. Such small mode volume, high $Q_o$ fiber cavities have revolutionized cavity QED experiments, especially the brightness of single photon sources \cite{najer2019gated} and we expect to see the same benefits for microwave to optical transduction experiments.  Using the $p_{11}$ coefficient of silica, this leads to an estimated $g_{0} \approx 2\pi*$ 6.5 kHz (at 7.33 GHz). The higher $g_{0}$ and $Q_o\approx5*10^4$ allow us to achieve a $C_{om} \approx$ 1 with an $N_{cav} \approx 3.36*10^6$, that can be accommodated at low temperatures without deleterious surface heating effects, potentially enabling continuous wave quantum transduction, even at dilution fridge temperatures.
    \item $Q_{m}$: The main unknown in estimating the performance of these devices is the $Q_m$ that shear mode devices can achieve at cryogenic temperatures, especially in amorphous materials like silica. In contrast to bulk mode HBARs which have been well studied at low temperatures and $Q_m > 10^6$ has been demonstrated at GHz frequencies, lateral field devices at these frequencies ($>$ 5 GHz) have been relatively less explored. In our estimation of $C_{om}$, we have used a conservative estimate of $Q_m \approx 5*10^4$ at cryogenic temperatures. In contrast to traditional HBARs where the acoustic wave is mostly confined under the metal electrode, shear mode HBARs should in principle exhibit higher $Q_m$ on account of being confined by dielectric boundaries and avoiding absorption losses in the metal electrodes.
\end{itemize}

\section{Conclusions}

We have demonstrated that MEMS-based suspended HBAR resonators are a viable cryogenic optomechanical platform for microwave to optical signal transduction. In addition the MEMS based scalable fabrication process allows the development of resonator arrays that can be used for fast free-space optical switching ($\sim$ 0.5 ${\mu}$s) in cryogenic environments. In contrast to 1D nanobeam optomechanical crystals, these devices provide a large phonon injection efficiency and large intracavity pump powers at cryogenic temperatures which can significantly enhance the transduction efficiency. By moving to lateral field excited shear mode resonators and building small mode volume optical cavities around these devices using fiber cavities, quantum transduction is within reach in this platform. 

\section{Acknowledgements}

KCB would like to acknowledge the European Research Council for funding support (ERC-StG, SBS 3-5, 758843). Nanofabrication work was carried out using equipment funded by the EPSRC capital grant, QuPIC (EP/N015126/1). We would like to thank Kartik Srinivasan, Seung-Bo Shim and Biswarup Guha for useful discussions and suggestions.

\appendix

\section{Optical cavity transfer matrix}\label{Ap:optical_spectrum}

The optical cavity spectra of the suspended HBAR resonators are modelled using a 1D transfer matrix \cite{hecht1998optics}. The devices are characterized by measuring the reflected optical signal off the device using the setup shown in Fig.\ref{fig:Setup}(a). The reflection spectrum can be modelled using a matrix product:

\begin{equation}
    M_{refl} = M_{I}M_{II} \dotsm M_{i} \dotsm M_{p} = \begin{bmatrix} m_{11} & m_{12} \\
    m_{21} & m_{22}\end{bmatrix}
\end{equation}

where
\begin{equation}
    M_{i} = \begin{bmatrix}
    \cos (k_i t_i) & \frac{i \sin (k_i t_i)}{Y_i}\\
    \frac{i \sin (k_i t_i)}{Y_i} & \cos (k_i t_i)
    \end{bmatrix}
\end{equation}

represents the transfer matrix of the layer $i$ in the stack, where $t_i$ is the film thickness, $k_i$ the wave vector and n$_i(\lambda_{opt})$ is the optical wavelength dependent refractive index of layer $i$, and the layer admittance ($Y_i$) is defined as:
\begin{equation}
Y_i = \sqrt{\frac{\epsilon_0}{\mu_0}}n_i
\end{equation}
with $\epsilon_0$ and $\mu_0$, the permittivity and permeability of the vacuum respectively. For normal incidence, which is the scenario in our experiments, the reflection coefficient ($r$) is defined as:
\begin{equation}
    r = \frac{Y_0 m_{11} + Y_0^2 m_{12} - m_{21} - Y_0 m_{22}}{Y_0 m_{11} + Y_0^2 m_{12} + m_{21} + Y_0 m_{22}}
    \label{eq:R_opt}
\end{equation}

\begin{figure}[!htbp]
    \centering
    \includegraphics[width = \columnwidth]{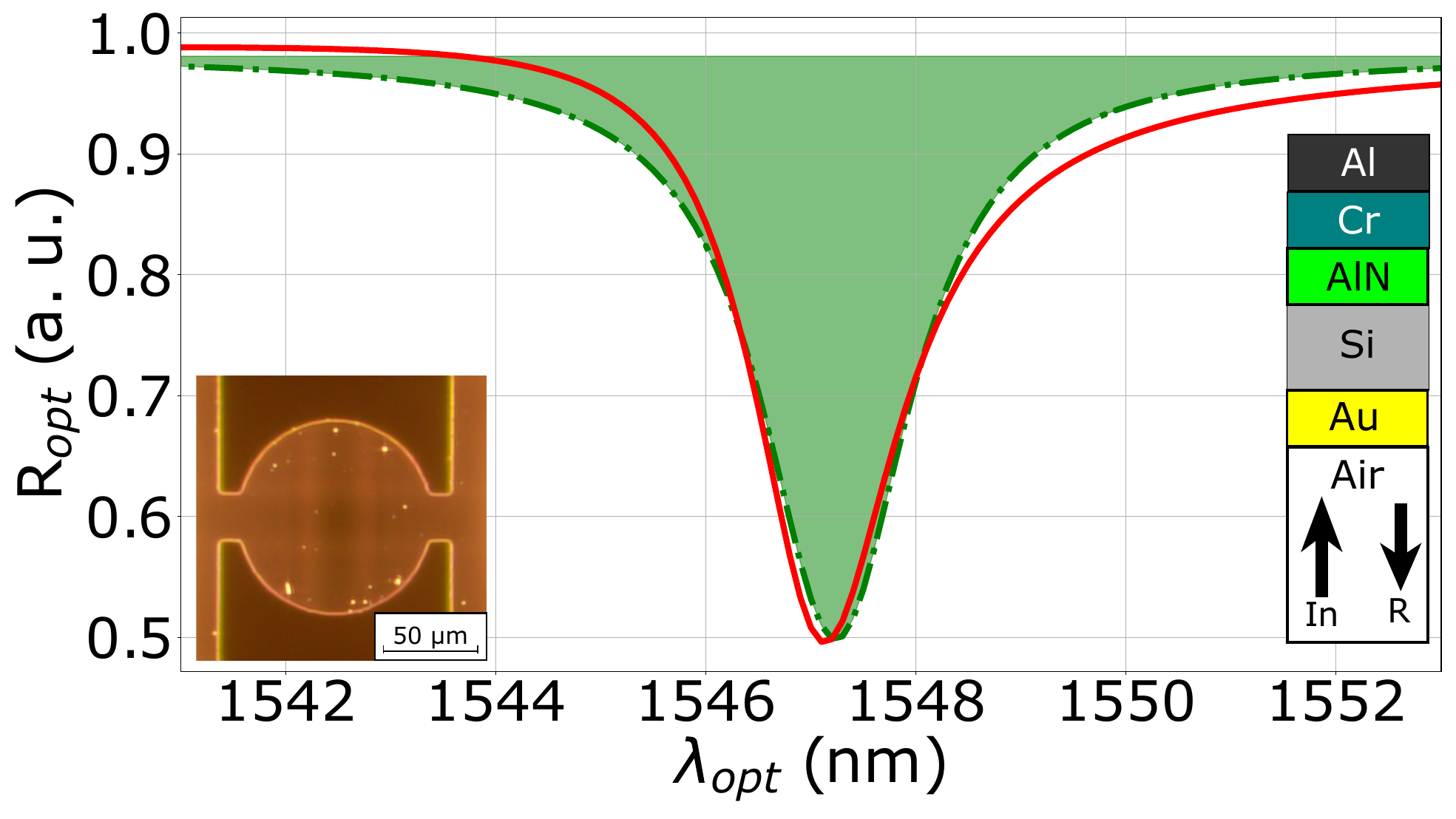}
    \caption{Simulated reflection spectrum (solid) and Lorentzian lineshape fit to extract Q$_{o}$(dashed) to the modified PiezoMUMPS layer stack: t$_{Al}$( 1 $\mu$m) - t$_{Cr}$( 20 nm) - t$_{AlN}$(0.5 $\mu$m) - t$_{Si}$( 10 $\mu$m) - t$_{Au}$ (23nm). A dark field image of the bottom face of the HBAR resonator with 23 nm gold deposited is shown in the inset.}
    \label{fig:Opt_trans}
\end{figure}

Fig. \ref{fig:Opt_trans} shows a representative simulated reflection spectrum of the PiezoMUMPS device stack.This model is used to estimate the thickness of the Au film deposited on the back side to help increase $Q_o$. Our model estimates $Q_o \approx$ 968 for $t_{Au} \approx$ 23 nm. In our experiments, we measure $Q_{o} \approx$ 460-500, with the difference mostly occurring due to surface roughness on the bottom silicon due to MEMS release processing (see inset of Fig. \ref{fig:Opt_trans}).

We also use the transfer matrix model to predict the temperature dependence of the optical cavity spectrum. In particular, we modify $n_{Si}(\lambda)$ to $n_{Si}(\lambda,T)$ using the measured temperature dependence of silicon \cite{Frey2006TemperaturedependentRI}. The model assumes that the refractive index of the other layers are temperature independent, but in practice, we find that we are able to achieve reliable fits to our experiments using just silicon's temperature dependence. This can be explained by the fact that the optical field predominantly resides in the silicon layer. The model also takes into account the overall change in cavity length due to cooling, but this contribution is negligible compared to the refractive index changes induced by temperature.

\section{The modulation index and photon transduction efficiency} \label{modulation index}

As described in detail in \cite{valle_high-frequency_2019}, the modulation index ($\beta$) is estimated from the modulated photocurrent signal measured using the setup shown in Fig.\ref{fig:Setup}(a). This signal represents the beat note between the optical pump and the (asymmetry) of the two sidebands generated by the AO modulation.

The output voltage at the photodetector can be represented as:

\begin{equation}
    V_{out} = \eta_{PD}Z_{load}\left| a_{out} \right|^2
\end{equation}
where $\eta_{PD}$ is the PD responsivity (A/W) and $Z_{load}$ the VNA input impedance (50 $\Omega$). $a_{out} = a_{in} - \sqrt{\eta_c \kappa}a_{cav}$ is the reflected field amplitude from the HBAR resonator, with $a_{in}$ the incident pump field. The intracavity field ($a_{cav}$) containing terms at the pump and the upper ($+\Omega_m$) and lower ($-\Omega_m$) mechanical sideband frequencies, in the limit of small modulation, is given by:

\begin{equation}
\begin{split}
    a_{cav} = & a_{in}\sqrt{\eta_c \kappa} \mathcal{L}(0) (R \\ &- \frac{i \beta \Omega_m \mathcal{L}(\Omega_m)}{2}e^{-i\Omega_m t} \\ & - \frac{i \beta \Omega_m \mathcal{L}(-\Omega_m)}{2}e^{+i\Omega_m t})
    \label{eq:a_cav}
\end{split}
\end{equation}
where $R$ is the cavity (field) reflectivity, $\eta_{c}$ is the cavity (field) coupling coefficient and $\mathcal{L}(\Omega)$ is the Lorentzian cavity response defined as:
\begin{equation}
\mathcal{L}(\Omega) = \frac{1}{\mathrm{-i}(\Delta + \Omega) + \kappa/2}
\label{eq:Lorentz}
\end{equation}
with $\Delta$ representing the pump detuning from the cavity resonance and $\kappa$ the cavity linewidth.

With some manipulation, $V_{out}$ can be expressed in terms of the cavity parameters as:
\begin{equation}
\begin{split}
    V_{out} = & 2 \Re \{ \mathrm{i} \eta_c \kappa P_{in} \Omega_m \frac{\beta}{2} ( R[\mathcal{L}_{S}^*\mathcal{L}_{0}^*-\mathcal{L}_{AS}\mathcal{L}_{0}] \\ & +\eta_c \kappa |\mathcal{L}_0|^2 \frac{\beta}{2}\left[\mathcal{L}_{AS}-\mathcal{L}_{S}^*\right] ) \} \eta_{PD} Z_{load}
    \label{eq:V_out_long}
\end{split}
\end{equation}
which can be inverted to extract $\beta$, given that the other parameters are independently known.

Knowing $\beta$, we can extract the optical power in the sidebands for a specific mechanical mode as a means of estimating the photon transduction efficiency:
\begin{equation}
    P_{sb} = \frac{\eta_c^2 \kappa^2 \mathcal{L}(0)^2 \beta^2 \Omega_m^2 \mathcal{L}^2(\Omega_m)}{2}P_{in}
\end{equation}

In the limit $\Delta\approx\kappa/2\gg\Omega_{m}$, this equation can be simplified as:
\begin{equation}
    P_{sb} \sim \left( {\frac{\eta_c\beta\Omega_m}{\kappa}} \right)^2P_{in}
\end{equation}

The CW photon transduction efficiency ($\eta_{peak}$) can then be extracted as:
\begin{equation}
    \eta_{peak} = \frac{P_{sb}}{\hbar\omega_{sb}}\frac{\hbar\Omega_{RF}}{P_{RF}}
\end{equation}

We can also express $\eta_{peak}$ in terms of the (electro) ($C_{em}$) optomechanical cooperativity ($C_{om}\ll$1)

\begin{equation}
    \eta_{peak} \approx \eta_{e}\eta_{o}4C_{em}C_{om}
\end{equation}

In our experiments, for the fundamental 340 MHz mode, $C_{om}\approx10^{-6}$ and $\eta_{peak}\approx10^{-7}$, which bounds $\eta_{e}\eta_{o}C_{em}< 10^{-1}$.

\section{Extraction of $Q_{m}$ using a multi-peak fit} \label{Lorentz Multi-peak fit}

The circular symmetry of the electrodes in the suspended HBAR devices leads to strong inter-mode coupling and the appearance of closely spaced peaks in the RF reflection ($S_{11}$) spectrum. The $Q_{m}$ of the resonator is therefore extracted from the S$_{11}$ spectrum (in Fig. \ref{fig:S11_HF}(b)) by a multi-peak Lorentzian fitting routine, where each Lorentzian resonance is defined as:

\begin{eqnarray}
    \mathcal{L}(\Omega_m) = \frac{A (w/2)^2}{(\Omega - \Omega_{m})^2+(w/2)^2}\\
\end{eqnarray}

A is an overall amplitude scaling parameter, $w$ is the full width at half maxima, $\Omega_m$ is the mode centre frequency. The $Q_{m}$ is then defined as:

\begin{equation}
    Q_{mech} = \frac{\Omega_m}{w}
\end{equation}

\begin{figure}[!htbp]
    \centering
    \includegraphics[width = \columnwidth]{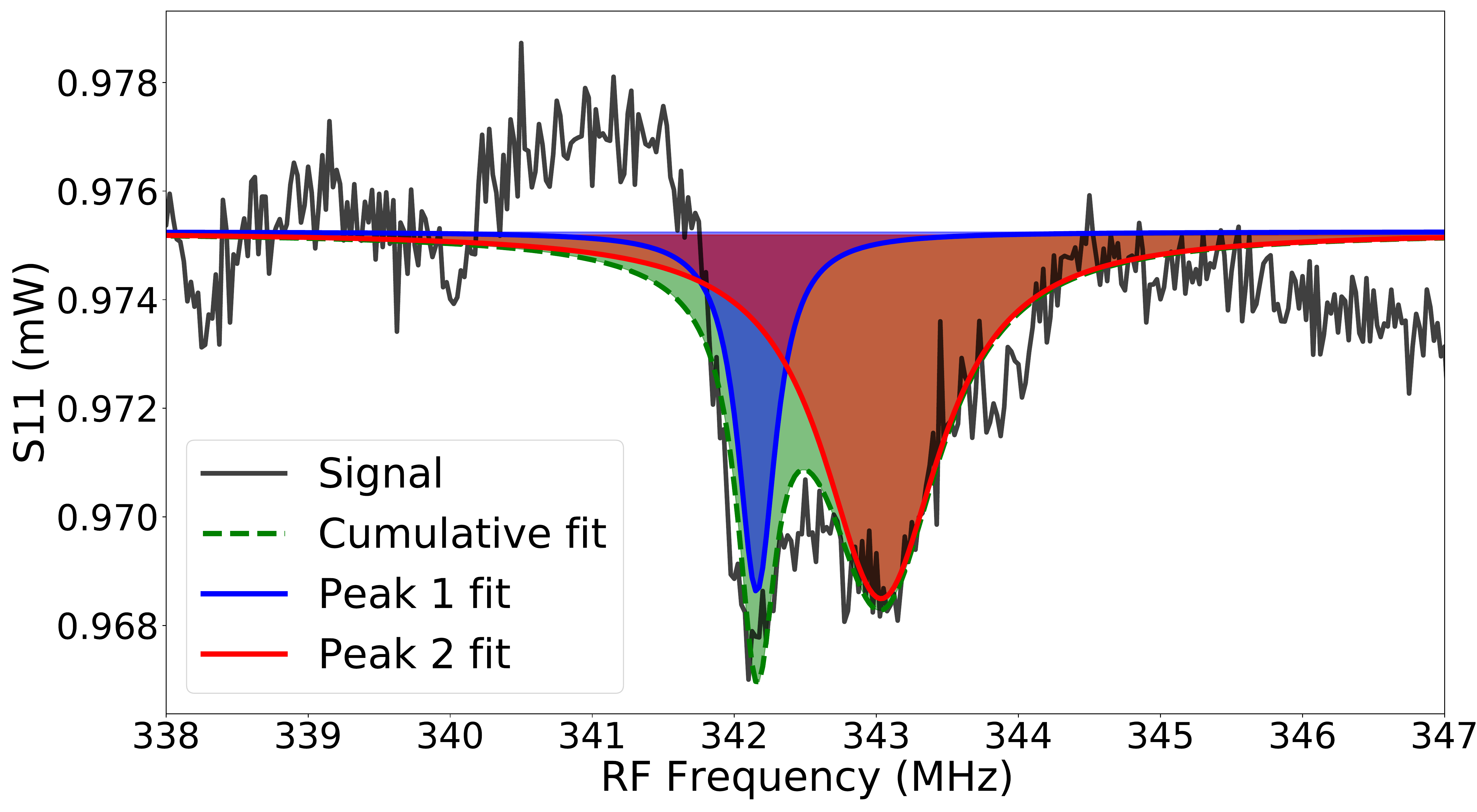}
    \caption{Multi-peak Lorentzian fit to the RF reflection ($S_{11}$) spectrum of the fundamental mechanical mode of the suspended HBAR device. We extract from the fit two closely spaced modes, with $f_c$ = 343.033 MHz; $Q_m$ = 1070 and $f_c$ 343.65 MHz; $Q_m$ = 339}
    \label{fig:MP-fit}
\end{figure}

When the $S_{11}$ spectrum shows closely spaced modes, a multi-peak ($\mathcal{L_{MP}}$) Lorentzian fit function is used where:

\begin{equation}
\mathcal{L}_{MP} = y_1 + \mathcal{L}_1(\Omega_m) + \dotsm + \mathcal{L}_N(\Omega_m) 
\end{equation}

where $y_{1}$ is the off-resonance $S_{11}$ magnitude. The $\mathcal{L}_{MP}$ allows us to estimate the $Q_{m}$ accurately. An example of the multi-peak fit routine applied to the fundamental ($\approx$ 342 MHz) mechanical mode is shown in Fig. \ref{fig:MP-fit}. The curve fitting is executed using a Python code which allows unsupervised analysis of data, to determine if a single peak (SP) or multi-peak (MP) fit to the data is needed.

\section{Fitting the fall time during pulsed operation of HBAR devices}
The fall time during cryogenic pulsed operation of the suspended HBAR devices is extracted from the oscilloscope traces by first normalizing the measured data (see Fig. \ref{fig:mixer_setup}(b)).

The normalized response is calculated as:
\begin{equation}
    V_{norm}= \frac{V_{meas} - V_{meas}^{min}}{V_{meas}^{max}-V_{meas}^{min}}
\end{equation}

We fit both the rise ($f_r$) and fall ($f_d$) times using exponential fits to extract the time constant ($\tau$) \cite{balram_acousto-optic_2017}:
\begin{equation}
f_r(t) = 1 - e^{-\frac{t}{\tau}}
\end{equation}
and
\begin{equation}
f_d(t) = e^{-\frac{t-t_{sh}}{\tau}}
\end{equation}
where $\tau$ is time constant, and t$_{sh}$ is the time shift applied to the function to fit the decay time since the trigger is on the positive edge (see Fig. \ref{fig:mixer_setup}(b)). As discussed in the text, the rise time shows evidence of ringing and overshoot at cryogenic temperatures that we don't observe at room temperature making the fits (and comparison) trickier, hence we use the fall time to quantify the response time of the system.

\section{Estimating $N_{cav}$} \label{N_cav}

\begin{figure}[!htbp]
    \centering
    \includegraphics[width = \columnwidth]{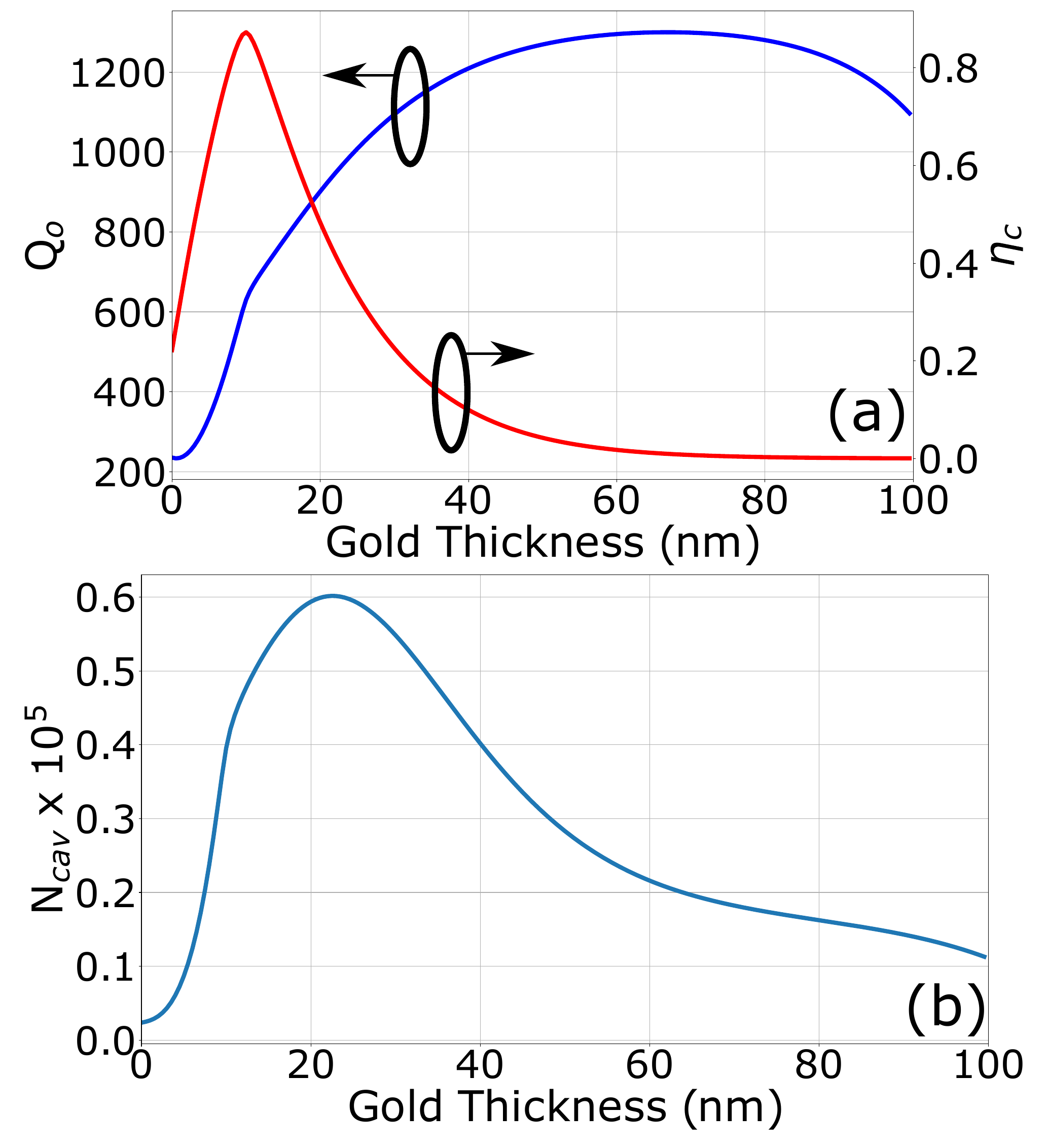}
    \caption{(a) Optical cavity quality factor ($Q_{o}$) and coupling coefficient ($\eta_c$) for the suspended HBAR devices with varying back mirror (Au) thickness (b) Steady state $N_{cav}$ plotted as a function of back metal thickness for fixed P$_{in}$ = 1 mW.}
    \label{fig:QvsAU}
\end{figure}

The steady state intracavity power ($P_{cav}$) for a given input power ($P_{in}$) is estimated as:

\begin{equation}
P_{cav} = \frac{\mathcal{F}}{\pi}P_{in}\frac{1-\mathcal{L}_{norm}(\Delta)}{1-\sqrt{(\mathcal{L}_{norm}(0)}}
\label{eq:P_cav}
\end{equation}

where $\mathcal{F}$ represents the cavity finesse and $\mathcal{L}(\Delta)$ the normalized Lorentzian cavity response at a detuning $\Delta$.

We can use $P_{cav}$ to estimate $N_{cav}$ at the optical pump frequency ($\omega_p$) as:

\begin{equation}
N_{cav} = P_{cav}\frac{\tau}{\hbar\omega_{p}}
\end{equation}
with the cavity photon lifetime defined as:
\begin{equation}
\tau = \frac{Q_{o}}{\omega_{c}}
\end{equation}

Since $C_{om} \propto g_{0}^2N_{cav}$, by using the model described in appendix \ref{Ap:optical_spectrum}, we can determine the optimal gold thickness to maximize $N_{cav}$. We estimate t$_{Au}$ $\approx$  23 nm. As can be seen in Fig. \ref{fig:QvsAU}(a), the optimal back mirror thickness represents a tradeoff between the $Q_{o}$ (increases with increasing $t_{Au}$) and $\eta_{c}$ (decreases with increasing $t_{Au}$).

\section{Optical power dependency} \label{Optical power dep}
\begin{figure}[!thbp]
    \centering
    \includegraphics[width = \columnwidth]{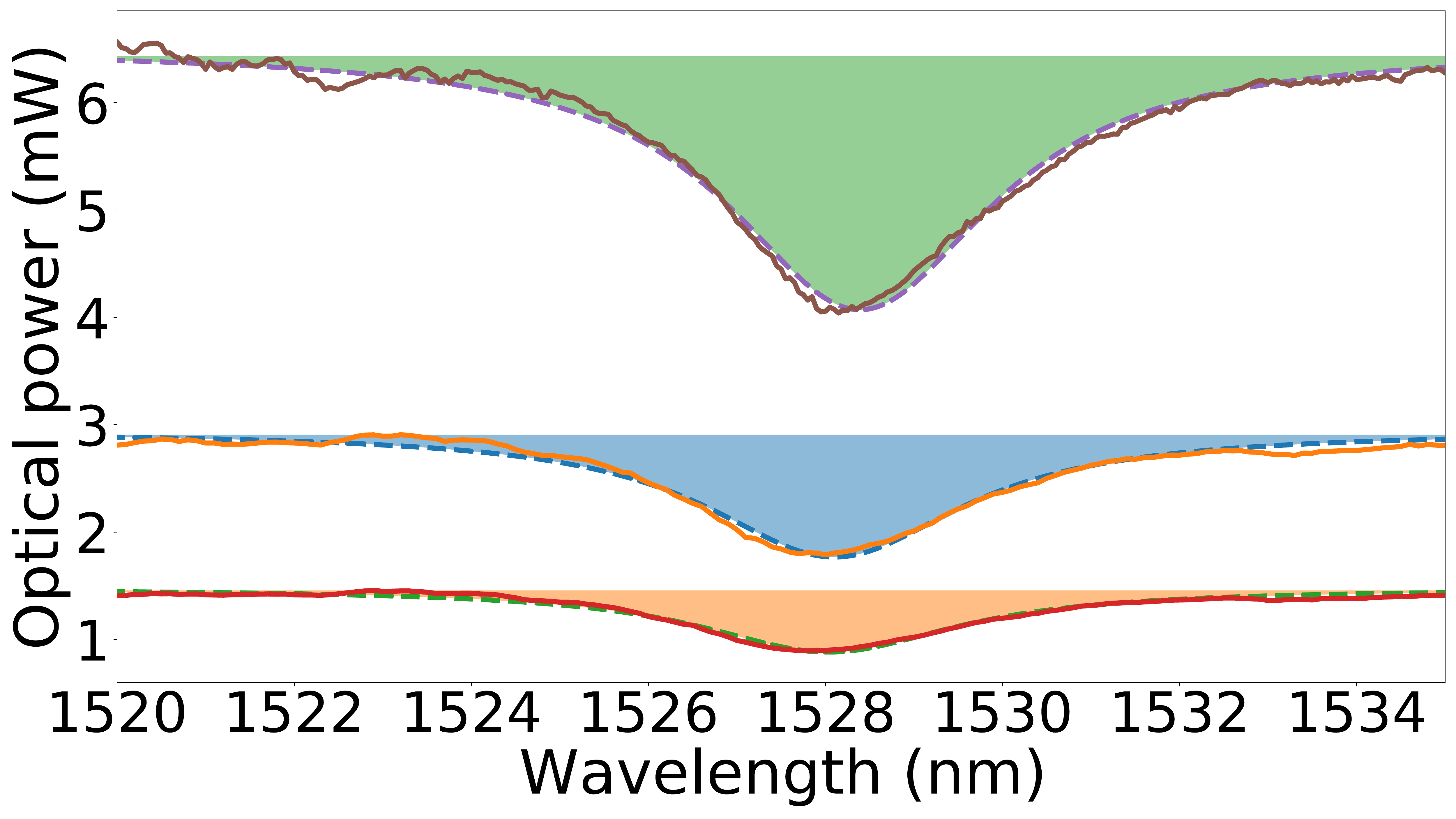}
    \caption{Measured optical cavity response at 10 K with varying $P_{in}$ to probe the effects of surface heating on the optical cavity response.}
    \label{fig:power_scan}
\end{figure}

One of the key advantages of the suspended HBAR platform for signal transduction is the ability to sustain a large $N_{cav}$ at cryogenic temepratures without suffering from deleterious heating effects \cite{balram2021piezoelectric}. To probe and understand surface heating, we measured the optical cavity response at 10 K for varying optical input powers ($P_{in}$). The results are shown in \ref{fig:power_scan}. Given the low $Q_{o}$ and the alignment dependence of the $Q_o$ for these devices, from the reported data, we cannot quantify the magnitude of the surface heating effects using our current experiments. We plan to replicate this experiment with the fiber based cavities proposed in Section \ref{quantum_hBAR}.\\

The higher $Q_{o}\approx 5*10^4$ achieved in the curved fiber cavities will provide a better evidence for the optical power handling capabilities of these devices.

\bibliography{piezo_om}
\end{document}